\documentclass[superscriptaddress,citeautoscript,prb,preprint]{revtex4-1}
\usepackage{graphicx}
\usepackage{enumitem}
\usepackage{gensymb}
\usepackage{amsmath}
\usepackage{amssymb}
\usepackage{dcolumn}
\usepackage{color}
\usepackage[multiple]{footmisc}
\begin{document}

\title{Double-layer ice from first principles}
\author{Ji Chen}
\email{ji.chen@ucl.ac.uk}
\affiliation{Department of Physics and Astronomy, London Centre for Nanotechnology, Thomas Young Centre, University College London, Gower Street, London WC1E 6BT, U.K.}
\author{Georg Schusteritsch}
\affiliation{Department of Materials Science and Metallurgy, University of Cambridge, 27 Charles Babbage Road, Cambridge CB3 0FS, U.K.}
\author{Chris J. Pickard}
\affiliation{Department of Materials Science and Metallurgy, University of Cambridge, 27 Charles Babbage Road, Cambridge CB3 0FS, U.K.}
\affiliation{Advanced Institute for Materials Research, Tohoku University 2-1-1 Katahira, Aoba, Sendai, 980-8577, Japan}
\author{Christoph G. Salzmann}
\affiliation{Department of Chemistry, University College London, 20 Gordon Street, London, WC1H 0AJ, U.K.}
\author{Angelos Michaelides}
\email{angelos.michaelides@ucl.ac.uk}
\affiliation{Department of Physics and Astronomy, London Centre for Nanotechnology, Thomas Young Centre, University College London, Gower Street, London WC1E 6BT, U.K.}

\begin{abstract}
The formation of monolayer and multilayer ice with a square lattice structure has recently been reported
on the basis of transmission electron microscopy experiments, renewing interest in confined two dimensional ice.
Here we report a systematic density functional theory study
of double-layer ice in nano-confinement.
A phase diagram as a function of confinement width and lateral pressure is presented.
Included in the phase diagram are honeycomb hexagonal, square-tube, hexagonal-close-packed and buckled-rhombic structures.
However, contrary to experimental observations, square structures do not feature: our most stable double-layer square structure
is predicted to be metastable.
This study provides general insight into the phase transitions of double-layer confined
ice and a fresh theoretical perspective on the stability of square ice in graphene nanocapillary
experiments.

\end{abstract}

%\maketitle must follow title, authors, abstract, \pacs, and \keywords
\maketitle

\section{introduction}

Recently, experimental evidence for the formation of two dimensional (2D) ice with a novel lattice structure in graphene nanocapillaries
was reported \cite{algara-siller_square_2015}.
Specifically, transmission electron microscopy (TEM) measurements showed that
2D ice appears as a layered structure with a square arrangement of oxygen atoms. 
Individual layers are located directly on
top of each other in an AA stacking manner, and
the thickest structures observed consisted of three layers.
However, the interpretation of the observations has been questioned and it has even been suggested
that the square lattice structure observed is not that of ice but rather sodium chloride, a
common contaminant \cite{zhou_observation_2015, algara-siller_algara-siller_2015, wang_wang_2015}. 
These observations and discussions have inspired great interest in further investigating 
confined 2D ice using advanced experimental and computational techniques.

\textit{Ab initio} methods such as density functional theory (DFT) and quantum Monte Carlo (QMC)
have recently been used to examine the stability of 2D ice structures.
These studies have significantly improved our understanding of 2D ice, particularly monolayer 2D ice \cite{chen_two_2016, corsetti_structural_2016, corsetti_enhanced_2016, roman_polymorphism_2016, chen_qmc}.
For example, in previous studies we have predicted a pentagonal monolayer 2D ice structure and
found that monolayer square ice is stable at high pressure, 
lending support to the measurements of Algara-Siller \textit{et al.} \cite{algara-siller_square_2015}
In addition, the benchmark-quality QMC data has shed light on the accuracy of force field (FF) models and DFT functionals \cite{chen_qmc}.
For instance, SPC/E \cite{berendsen_missing_1987} and TIP4P \cite{jorgensen_comparison_1983} type models were found to overbind high density 2D ice phases.
Meanwhile, several exchange-correlation (XC) functionals, in particular the rPW86-vdW2 \cite{lee_higher-accuracy_2010}, were identified that predict relatively correct binding energies for both 2D and 3D ice.
This suggests that DFT can be used in further investigations of confined ice beyond the monolayer.

Apart from monolayer 2D ice, double-layer ice is of great importance
to the questions surrounding square ice as it was also observed in the experiments of Algara-Siller \textit{et al.}.
However, due to the possibility of forming interlayer hydrogen bonds a simple extension of monolayer ice phase diagram to
the double-layer would be inappropriate.
%As part of the experimental study FF simulations were also performed.
%These identified a monolayer square ice structure but failed to explain the AA stacking order of 
%the double-layer square ice \cite{algara-siller_square_2015}.
Indeed, double-layer ice has been discussed in many FF studies,
yet observations were quite sensitive to the FF model and computational setup used \cite{algara-siller_square_2015, koga_freezing_1997, koga_first-order_2000, zangi_bilayer_2003, giovambattista_phase_2009, han_phase_2010, johnston_liquid_2010, kastelowitz_anomalously_2010, bai_polymorphism_2012, giovambattista_computational_2012, zhao_highly_2014, lu_investigating_2014,  Mario_AA_2015, corsetti_enhanced_2016, Zubeltzu_pre}.
While double-layer ice with the AA stacking
observed in experiments has been formed in
some simulations \cite{algara-siller_square_2015, Mario_AA_2015},
many other studies have reported different structures.
%Simulations using the TIP4P water model by 
%Koga, Tanaka and Zeng first observed the freezing of water into double-layer hexagonal crystalline and 
%amorphous phases \cite{koga_freezing_1997, koga_first-order_2000}.
%Later, 
%using the TIP5P water model Zangi and Mark reported a high density rhombic phase \cite{zangi_bilayer_2003}.
%Bai and Zeng observed the hexagonal, the rhombic and another high density phase that resembles a parallel arrangement of 1D square tubes \cite{bai_polymorphism_2012}.
%Johnston, Kastelowitz and Molinero reported flat hexagonal, buckled hexagonal,
%pentagonal, quasicrystal and amorphous phases based on simulations 
%using the coarse-grained mW water model \cite{johnston_liquid_2010, kastelowitz_anomalously_2010}.
%Han \textit{et al.}
%found that at low densities hexagonal ice forms whereas at high densities the hexatic liquid crystal phase was observed.
%A similar structure, which was dubbed as a triangular phase, was also identified by Corsetti \textit{et al.} \cite{corsetti_enhanced_2016}
Bearing in mind that these FFs encounter problems for monolayer ice \cite{chen_qmc}, question marks have to be raised regarding their performance for double-layer ice.
Therefore, a systematic DFT study of double-layer ice using reliable DFT XC functionals is needed.
In particular the stacking order and the interlayer interaction in 2D ice are yet to be investigated.
With this in mind, herein we report a DFT study aimed at exploring double-layer ice phases under confinement.
The aims are to
establish the stability of various
double-layer ice structures as a function of lateral pressure and width of confinement, 
and to shed light on the experimental 
observation of AA stacking square ice.

The remainder of this paper is organised as follows. 
Section II describes the details of various DFT calculations.
Section III reports and discusses the main results, wherein: 
(i) we propose a phase diagram for double-layer ice at 0 K as a function of confinement width 
and pressure up to 5 GPa;
and (ii) we discuss the stability of double-layer square ice.
In section IV we provide a brief summary of our results.

\section{simulation details}
We have carried out \textit{ab initio} random structure searches (AIRSS) \cite{pickard_ab_2011} using two different schemes:
(i) A fully random structure search with 8, 12, and 24 water molecules per unit cell;
and (ii) A limited random structure search starting with hexagonal, pentagonal, square and HCP lattices 
containing randomly orientated water molecules. 
For the latter approach 12 water molecules were considered for the hexagonal unit cell, 24 for the
pentagonal cell, 8 and 18 for the square and HCP lattices, respectively.
Periodic boundary conditions were used with two layers of ice and a vacuum region outside the ice layers.
K-point sampling was used with Monkhorst-Pack grids with 
a lateral separation between points larger than 0.03 $\text{\AA}^{-1}$.
The lateral cell dimensions were relaxed until the lateral stress tensor
converged to the target lateral pressures (0 - 5 GPa).
The cell dimension perpendicular to the slab was fixed during structure optimization.
DFT calculations were performed using the Vienna \textit{Ab initio}
Simulation Package (VASP) where the core electrons were described with projector 
augmented wave (PAW) potentials \cite{kresse_efficient_1996, kresse_ultrasoft_1999}.
An energy cut-off of 550 eV was used in the structure search and phase diagram calculations.
The absolute binding energy values referred to in the text and reported in the tables are 
calculated using hard PAW potentials in conjunction 
with a 1000 eV cut-off.
The bulk of the results are based on
a non-local van der Waals inclusive XC functional 
rPW86-vdW2 (often known as vdW-DF2) \cite{lee_higher-accuracy_2010}.
This functional, as implemented in VASP by Klime\v{s} \textit{et al.} \cite{klimes_van_2011}, has proved 
to be suitable for predicting the binding of monolayer and bulk ice phases \cite{santra_accuracy_2013, chen_qmc}. 
Nevertheless, the optPBE-vdW and optB88-vdW functionals \cite{klimes_chemical_2010, klimes_van_2011},
the Perdew-Burke-Ernzerhof (PBE) functional \cite{perdew_generalized_1996} and 
the Heyd-Scuseria-Ernzerhof (HSE) hybrid functional \cite{heyd_hybrid_2003, heyd_erratum:_2006} with 
the van der Waals correction of Tkatchenko and 
Scheffler \cite{tkatchenko_accurate_2009} (PBE+vdW(TS), HSE+vdW(TS)) 
have also been tested for some specific structures.

The confinement was not modelled by explicit graphene sheets but rather with 
a uniform 2D confining potential. Specifically a Morse potential was fitted to QMC
results for the interaction of a water monomer with graphene \cite{ma_adsorption_2011}.
The potential $V(z) = D((1-e^{-a(z-z_0)})^2-1) $, where z is the distance between the oxygen atom and
the wall, $D = 57.8 $ meV, $a = 0.92~\text{\AA}^{-1}$, $z_0 = 3.85~ \text{\AA}$.
More details on this confining potential can be found in Ref. \onlinecite{chen_two_2016} and
the supplemental material \cite{si}.
Calculations of 2D ice confined within actual sheets of graphene have also been performed to estimate the
optimal separation of graphene layers as discussed below.

The binding energy is defined as,
\begin{equation} \label{eq1}
E_\text{b}=E^\text{tot}_{\text{H}_\text{2}\text{O}} - E^\text{tot}_\text{ice}/n_{\text{H}_\text{2}\text{O}},
\end{equation}
where
$E^\text{tot}_{\text{H}_\text{2}\text{O}}$ is the total energy of a water molecule in vacuum, 
$E^\text{tot}_\text{ice}$ and $n_{\text{H}_\text{2}\text{O}}$
are the total energy and number of water molecules in the 2D ice structures.
The enthalpy is defined as,
\begin{equation} \label{eq2}
H = E^\text{tot}_\text{ice} + E_\text{confinement} + P \times A \times h,
\end{equation}
where $E_\text{confinement}$
is the energy in the confinement potential,
$A$ is the lateral area,
$h$ is the layer height which equals the width of the confinement $w$,
and $P$ is the lateral pressure \footnotemark[1].

Some short \textit{ab initio} molecular dynamics (AIMD) simulations were also carried out 
with a view to understand the possible role of anharmonic effects.
These simulations were performed within the canonical ensemble at a target temperature of
300 K using a Nos\'{e}-Hoover thermostat \cite{nose_unified_1984}.
Each simulation was performed with a time-step of 0.5 fs for a total of 10 ps.
Of these 10 ps, the first 3 ps were used for equilibration and analysis was performed on the remaining 7 ps.
The phonon densities of states were also calculated
through Fourier transformation of the velocity auto-correlation function from
these molecular dynamics trajectories.

\section{Results and discussions}

\subsection{Structures and binding energies of double-layer ice at ambient pressure}

\begin{figure}[h!]
\begin{center}
\includegraphics[width=8.5cm]{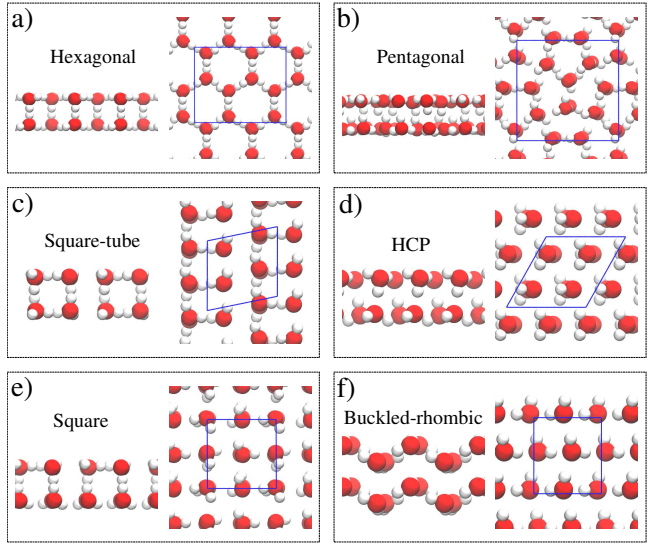}
\end{center}
\caption{Double-layer ice structures.
Side and top views of the (a) hexagonal (honeycomb), (b) pentagonal,
(c) square-tube, (d) hexagonal-close-packed (HCP),
(e) square
and (f) buckled-rhombic double-layer structures.
Red and white spheres represent oxygen and hydrogen atoms, respectively.
The blue boxes in the top views show the unit cells used in the periodic DFT calculations.
All the structure files are provided in the supplemental material \cite{si}.
}
\label{figure1}
\end{figure}

We start our discussion by looking at the most relevant double-layer ice structures 
and their binding energies at ambient pressure (Fig. \ref{figure1}, Table I).
They have initially been obtained with a confinement of 9.5 \AA~and re-optimized
after the confinement is removed.
The most stable is a hexagonal structure, which consists
of two layers of 2D honeycomb lattice that 
locate directly on top of on another with AA stacking (Fig. \ref{figure1}a).
The hexagonal structure has been 
observed quite often in force field studies \cite{koga_freezing_1997, witek_structure_1999, han_phase_2010, kastelowitz_anomalously_2010, johnston_liquid_2010, bai_polymorphism_2012}, 
and has also been suggested experimentally
on a metal supported graphene surface \cite{kimmel_no_2009}.
Second most stable is a double-layer pentagonal structure (Fig. \ref{figure1}b), reminiscent of a Cairo tiling pattern.
This structure was first observed by Johnston \textit{et al.} through quenching of liquid water using both 
the mW and TIP4P/ice water models \cite{johnston_liquid_2010}.
Similarly, a monolayer pentagonal structure was predicted in our previous DFT study \cite{chen_two_2016}.
The monolayer pentagonal ice has a vanishing energy 
difference 
to the monolayer hexagonal phase \cite{chen_qmc}.
Surprisingly, we find that the double-layer pentagonal structure 
is 25 meV/$\text{H}_\text{2}\text{O}$ less stable than the hexagonal double-layer.
Following these two we have 
the square-tube phase (Fig. \ref{figure1}c),
the hexagonal-close-packed (HCP) phase (Fig. \ref{figure1}d), 
and the double-layer square structure (Fig. \ref{figure1}e).
The double-layer square structure is a new structure identified in the current AIRSS study.
As with the hexagonal, pentagonal, square-tube, and HCP structures the double-layer square structure is held together with
interlayer hydrogen bonds in an AA stacking arrangement. 
This double-layer AA stacked square ice structure is a candidate for the 
AA stacked structure observed in experiments \cite{algara-siller_square_2015}.
It is the most stable double-layer square ice structure identified in our structure searches using
both the 8 and 18 water molecule unit cells.
It has a binding energy of 511 meV/$\text{H}_\text{2}\text{O}$,
being less stable than the most stable
hexagonal structure by 37 meV/$\text{H}_\text{2}\text{O}$.
Apart from the above structures, Fig. \ref{figure1}f also shows a double-layer buckled-rhombic structure.
The interlayer interaction in this structure is mediated by van der Waals forces and it is free of
interlayer hydrogen bonds.
The buckled-rhombic phase is only stable under confinement. 
We will have more to say about this structure later.
We note that all the structures identified here are non-polar, and in this study the 
potential impact of an external electric field
is not considered \cite{zhao_ferroelectric_2014, fernandez_electric_2016}.

\begin{table}[h]
\resizebox{0.45\textwidth}{!}{
 \begin{tabular}{|c|c|c|c|}
  \hline
  Structure   & $E_b$ (meV/$\text{H}_\text{2}\text{O}$)   & $A$ (\AA$^2$/$\text{H}_\text{2}\text{O}$) & $d$ (\AA)\\
  \hline
  Hexagonal    & 548/575   & 10.10/9.76  & 2.85/2.80    \\
  Pentagonal   & 523/553   & 8.88/8.65   & 2.86/2.81    \\
  Square-tube         & 523/549   & 8.27/8.14   & 2.85/2.80    \\
  HCP          & 520/550   & 8.07/8.10   & 2.75/2.68    \\
  Square       & 511/538   & 8.49/8.40   & 2.86/2.80    \\
  \hline
 \end{tabular}
}
 \caption{
Binding energies (Eq. \ref{eq1}) and structural information of
double-layer ice structures at ambient pressure without confinement. 
Larger values for $E_b$ suggest stronger binding.
A is the lateral area per water molecule within a single layer.
d is the average interlayer distance.
Hexagonal represents the hexagonal honeycomb structure (Fig. \ref{figure1}a) 
and HCP represents the hexagonal-close-packed structure (Fig. \ref{figure1}d).
Results using both the rPW86-vdW2 and optPBE-vdW functionals are reported (separated 
by a slash with rPW86-vdW2 in front).
}
\label{table1}
\end{table}

%\textcolor{red}{phase transition at 9.5 \AA, graphene condition}

\subsection{Phase transitions of double-layer ice}

\begin{figure}[h!]
\begin{center}
\includegraphics[width=7cm]{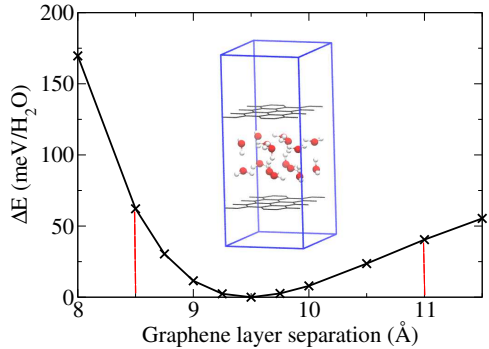}
\end{center}
\caption{Total energy profile of water confined within
two perfectly flat sheets of graphene fixed at the separations shown.
The dashed lines show the confinement width regime examined in this study.
The inset shows the structural model used for this specific set of calculations, which involves 16 water molecules 
in a hexagonal double-layer ice structure confined between two 32 carbon atom graphene sheets.
On the energy scale the energy minimum is set to zero.
}
\label{figure2}
\end{figure}

\begin{figure}[h!]
\begin{center}
\includegraphics[width=8.5cm]{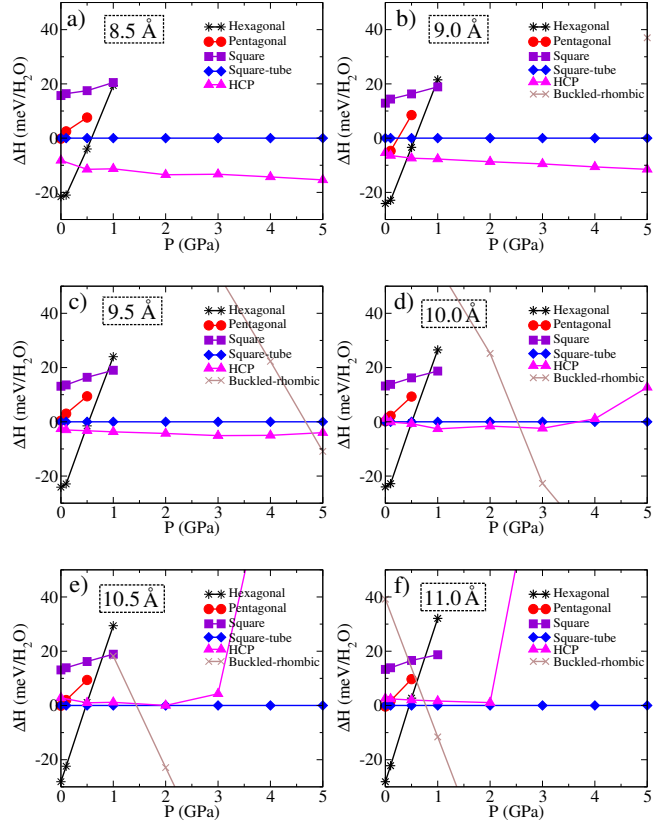}
\end{center}
\caption{Enthalpies (relative to square-tube) of the double-layer ice structures 
as a function of lateral pressure
for different confinement widths.
These are from (a) to (f): 8.5, 9.0, 9.5, 10.0, 10.5 and 11.0 \AA.
The enthalpies of the hexagonal and square phases and the pentagonal phase are only
shown up to 1 GPa and 0.5 GPa, respectively, because
they are structurally unstable above this. 
The unstable structures and additional data showing the instability of these phases is included in the supplemental material. 
}
\label{figure3}
\end{figure}

Examining the phase transitions at different confinement widths is desirable as the
phase stability of confined ice is sensitive to the width of confinement \cite{zangi_monolayer_2003, zangi_bilayer_2003, giovambattista_phase_2009, kastelowitz_anomalously_2010, chen_two_2016}.
However, since a confinement that mimics graphene nanocapillaries is particularly interesting
we first established what the appropriate confinement width for graphene is 
by examining the total energy of a double-layer ice film within two 
layers of graphene (Fig. \ref{figure2}). 
These calculations show that the optimal width of confinement is around 9.5 \AA.
This is the main reason our structure searches were performed at 9.5 \AA~confinement.
However
we also examined the stability of the relevant ice structures at a broader range of confinement.
Fig. \ref{figure3} shows the relative enthalpy with respect to the square-tube structure for confinements between 8.5 and 11.0 \AA.
At confinement widths of 8.5 and 9.0 \AA, just a single transition is observed from the hexagonal to the HCP structure at
about 0.5 GPa.
At 9.5 \AA~the hexagonal structure is still the most stable structure below \textit{ca.} 0.5 GPa.
However, the square-tube structure becomes more stable than the HCP structure, leading to a new phase transition
from the hexagonal to the square-tube structure.
The HCP structure is slightly more stable than the square-tube at pressures above \textit{ca.} 1.5 GPa and transforms
to the buckled-rhombic structure at \textit{ca.} 5 GPa.
The transition between these two high density double-layer ice phases has also been identified in
Ref. \onlinecite{corsetti_enhanced_2016}.
As the width of confinement increases: 
(i) the stability of the hexagonal structure is not altered; 
(ii) the HCP and square-tube structures have very similar stability in a wide range of pressures, within which 
the square-tube structure is slightly preferred at wider confinements;
and (iii) the transition pressure to the buckled-rhombic structure decreases, 
squeezing the stable region of the square-tube and HCP structures.
Generally, the width dependency is strong for the HCP and buckled-rhombic structure whereas the relative
stability of the hexagonal, pentagonal, square and square-tube phases barely changes at different confinement widths.
This is not so surprising as the hexagonal, pentagonal, square and square-tube structures have the same
number of interlayer hydrogen bonds and similar interlayer separations.
The HCP structure is thinner and would be favored at narrow confinement
whilst the buckled-rhombic structure is thicker and tends to appear in wider confinements.

\footnotetext[1]{
$P=\frac{1}{2}(\sigma_{xx}+\sigma_{yy})$,
$\sigma=\sigma'\times L_z /h$,
$\sigma'_{xx}$ and $\sigma'_{yy}$
are the calculated lateral (x and y direction) diagonal stress tensor elements for the slab/vacuum model,
$L_z$ is the length of the cell in the out-of-plane direction.
In the calculation $\sigma'_{xx}\times L_z$ and $\sigma'_{yy}\times L_z$ are 
conserved quantities, the calculation of the enthalpy is independent 
of the definition of $h$. Therefore, the shape of the phase diagram is not affected by the definition.
However, the definition of $h$ does affect the values of the transition pressures predicted.
Our definition indicates an upper limit of layer height (width of confinement), corresponding to a lower limit of
transition pressures. Assuming the lower limit of width is $w - 3.0$ ($w$ = 9.5) 
the upper limit of the transition pressures are about 1.5 times larger.
}

%\textcolor{red}{change the width, 2d phase diagram}

\begin{figure}[h!]
\begin{center}
\includegraphics[width=8cm]{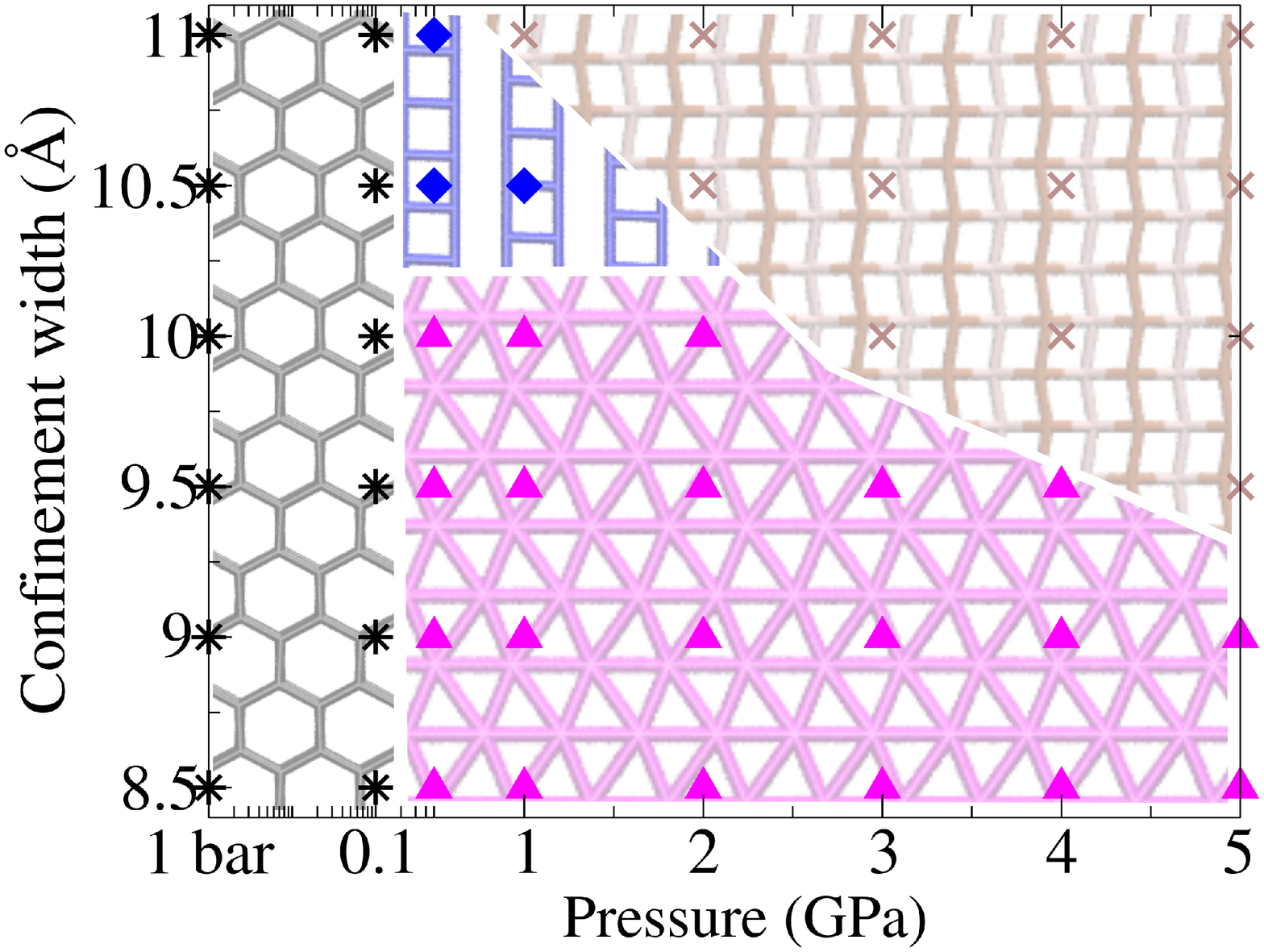}
\end{center}
\caption{
Schematic
phase diagram of double-layer ice with respect to lateral pressure and confinement width.
The phase diagram can be divided into four regions:
(i) The hexagonal phase at low pressures;
(ii) The HCP phase at high pressures and small confinement widths;
(iii) The square-tube phase at pressures in the GPa regime and large confinement widths;
(iv) The buckled-rhombic phase at high pressures and large confinement widths.
The specific data points calculated are shown with the same set of symbols as in Fig. \ref{figure3}.
Note that a logarithmic scale is used for pressure from 1 bar to 0.5 GPa 
and then a linear scale from 0.5 GPa to 5 GPa.
We note that the enthalpy difference between the HCP and square-tube structures is so small that the 
boundary between these two phases is not well defined from our calculations, 
from which we see a region (0.5 - 1 GPa, 9.5 - 10 \AA) that the hexagonal, HCP and square-tube structures are intermixed.
However, due to the fact that the HCP has smaller interlayer distance and higher lateral density than the square-tube,
the transition boundary between them is more likely to have a positive slope as shown in the figure.
}
\label{figure4}
\end{figure}

The sequence of phase transitions at different confinement widths allows us
to sketch a putative phase diagram for double-layer ice at 0 K as a function of pressure and confinement 
width (Fig. \ref{figure4}).
In brief, double-layer ice appears as a hexagonal structure at low pressures ($\sim$ 0.5 GPa).
This then transforms to the HCP 
structure at higher pressures for narrow confinement widths
and to the square-tube structure for larger confinement widths.
At higher pressures and larger confinement widths double-layer ice 
favors a buckled-rhombic structure.

\subsection{Stability of double-layer square ice}
%\textcolor{red}{compare with experiment, square ice}

Contrary to the monolayer, our predicted phase diagram for double-layer ice does not have a
regime in which square ice is stable, whereas it is 
the only stable ice structure identified so far that matches the TEM images 
in graphene nanocapillaries \footnotemark[2].
Our calculations reveal that irrespective of the confinement width the square structure is always at least \textit{ca.} 20 
meV/$\text{H}_\text{2}\text{O}$ less stable than the most stable structure (Fig. \ref{figure3}).
This is consistent with another DFT study in which slightly different computational settings were
used and double-layer square ice was also not identified as 
a stable phase \cite{corsetti_enhanced_2016}.
Intuitively the fact that the double-layer square phase is metastable is not so surprising
by just comparing the square phase with the square-tube phase.
Two steps can turn a double-layer square structure to a square-tube structure: changing the hydrogen ordering of the square phase and shifting
the tubes by half a lattice unit.
The re-ordering of hydrogen bonds is unlikely to cause a significant energy increase whereas the tube formation allows a further optimization step towards a more stable structure, the square-tube structure.
However, due to the small energy differences between the different phases 
further calculations are needed to test this conclusion.
%There could be many reasons why our calculations do not identify the square structure
%as a stable phase in the phase diagram of double-layer ice.
%There could be related to the specific conditions of the experiments,
%kinetic factors or the accuracy of the simulations themselves.
Naturally it is reasonable to first consider the limitations of
the simulations and to this end we specifically consider the following issues:
(i) The accuracy of the underlying DFT calculations;
(ii) The role of zero point energies;
(iii) Finite temperature effects including harmonic and anharmonic phonon contributions;
and (iv) Hydrogen ordering and the influence of configurational entropy.
Each of these issues is addressed at a confinement width of 9.5 \AA.

\begin{figure}[h!]
\begin{center}
\includegraphics[width=7cm]{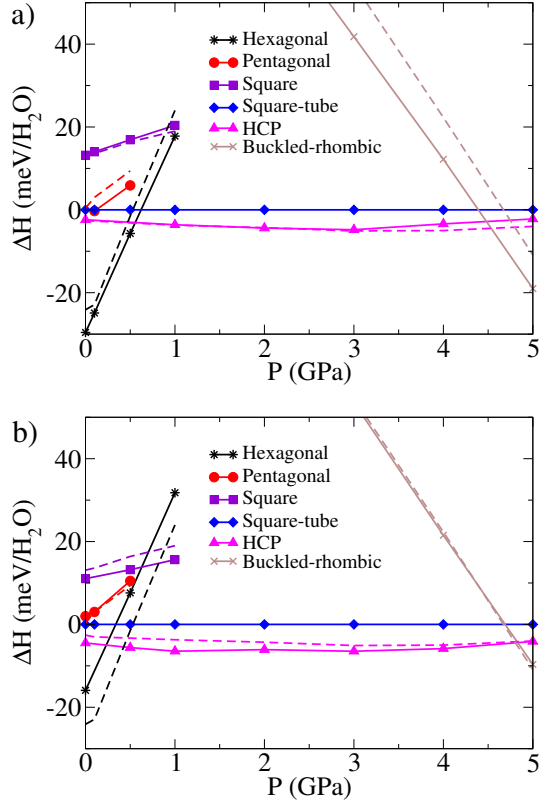}
\end{center}
\caption{
(a) Comparison of the relative enthalpies obtained with
the optPBE-vdW (solid line) and rPW86-vdW2 (dashed lines) functionals.
(b) Comparison of the relative enthalpies obtained from rPW86-vdW2 
with (solid line) and without (dashed lines) zero point energy (ZPE).
ZPEs have been calculated as,
$E_\text{ZPE}=\sum_{i=1}^{i=N} \frac{1}{2} \hbar \omega_{\Gamma,i}$
where $\omega_{\Gamma,i}$ is the $i$th gamma point vibrational frequency
calculated using the finite displacement method, and $\hbar$ is the reduced Planck constant, and N
is the number of vibrational modes.
}
\label{figure5}
\end{figure}

\textbf{Sensitivity to exchange correlation functional:}
The first issue is the accuracy of DFT calculations
specifically the XC functional used.
The results reported so far have been obtained with the rPW86-vdW2 functional, which has proved to be accurate
in predicting the binding energies of monolayer and bulk ice polymorphs \cite{santra_accuracy_2013, chen_qmc}.
Nevertheless, it is also worth  examining the results with other functionals.
In Table \ref{table1} and  Fig. \ref{figure5}a we report the binding energy, lattice parameter and enthalpy 
calculated using
the optPBE-vdW functional.
% which has been shown to give better densities of bulk ice phases \cite{santra_accuracy_2013}.
As shown in Fig. \ref{figure5}a, although the two functionals differ in their prediction of absolute binding energies and lattice parameters,
the relative enthalpies between different phases mostly agree.
Importantly, the difference between the square structure and the stable structures are consistent 
for these two functionals.
Specifically at 1 GPa, the pressure regime at which square ice is estimated to 
form \cite{algara-siller_square_2015}, the square-tube structure has a
lower enthalpy by \textit{ca.} 19 meV/$\text{H}_\text{2}\text{O}$ than the square structure.

There are, of course, many other XC functionals we could consider \cite{gillan_perspective}.
Focusing on 1 GPa we have explored how several other functionals perform,
including those with a different treatment of van der Waals and a hybrid exact exchange functional.
The 
key results are given in Table \ref{table2}.
Overall we find that the difference between the square-tube and square structure is not
very sensitive to the choice of XC functional and for all functionals square-tube is 
\textit{ca.} 20 meV/$\text{H}_\text{2}\text{O}$ more stable than the square structure at 1 GPa.
The apparent insensitivity to the XC functional can be further shown by the decomposition of
enthalpy in Table \ref{table2}.
%In order to understand this apparent insensitivity to the XC functional
%we decomposed the enthalpy differences into the water binding energy within the ice structures, the confinement energy and
%the volume-pressure term in Table \ref{table2}.
%This decomposition allows us to understand the origin of the 
%enthalpy difference between these two structures.
%First, the square and square-tube structures have similar interlayer distances, 
%thus confinement contributes equally.
%Second, the square-tube structure has a larger binding energy and smaller density than the square structure that three quarters of the
%total enthalpy difference comes from the binding energy difference while the rest
%results from the difference in density.
%This explains why different methods give such similar results for the difference between these two structures
%and indicates that the conclusion is largely insensitive to the XC functional used.

\begin{table}[h]
\resizebox{0.45\textwidth}{!}{
 \begin{tabular}{|c|c|c|c|c|}
  \hline
  Methods   & $\Delta H$   & $\Delta E_\text{ice}^\text{tot}$  & $\Delta E_\text{confinement}$  & $\Delta (\text{P} \times \text{A} \times \text{h}) $\\
  \hline
  rPW86-vdW2 & 19  & 14 & 0   & 5    \\
  optPBE-vdW & 20  & 15 & 0   & 5    \\
  optB88-vdW & 24  & 18 & 0   & 6    \\
  PBE+vdW(TS)  & 23  & 16 & 0   & 7    \\
  HSE+vdW(TS) & 25  & 19 & 0   & 6    \\
  \hline
 \end{tabular}
}
 \caption{
Stability of the square-tube and double-layer square structures at 1 GPa calculated using different XC
functionals at a confinement width of 9.5 \AA.
$\Delta H$, $\Delta E_\text{ice}^\text{tot}$, $\Delta E_\text{confinement}$ 
and $\Delta (P \times A \times h)$ are differences
between the square and the square-tube structure for energy terms as defined in 
Eqs. \ref{eq1} and \ref{eq2}.
All values are reported in units of meV/$\text{H}_\text{2}\text{O}$.
}
\label{table2}
\end{table}

\textbf{Role of zero point energies:}
Since we consider small energy differences between the various phases
it is possible that differences in zero point energy (ZPE) could tip the balance 
in stability between them.
Therefore we computed 
the ZPE of relevant phases within the harmonic approximation.
As shown in Fig. \ref{figure5}b no
significant changes in phase transitions have been identified when ZPEs are accounted for.
Specifically, comparing the enthalpy of the square structure and the square-tube structure at 1 GPa,
the relative stability of the square structure is slightly 
increased by \textit{ca.} 4 meV/$\text{H}_\text{2}\text{O}$.

\begin{figure}[h!]
\begin{center}
\includegraphics[width=7cm]{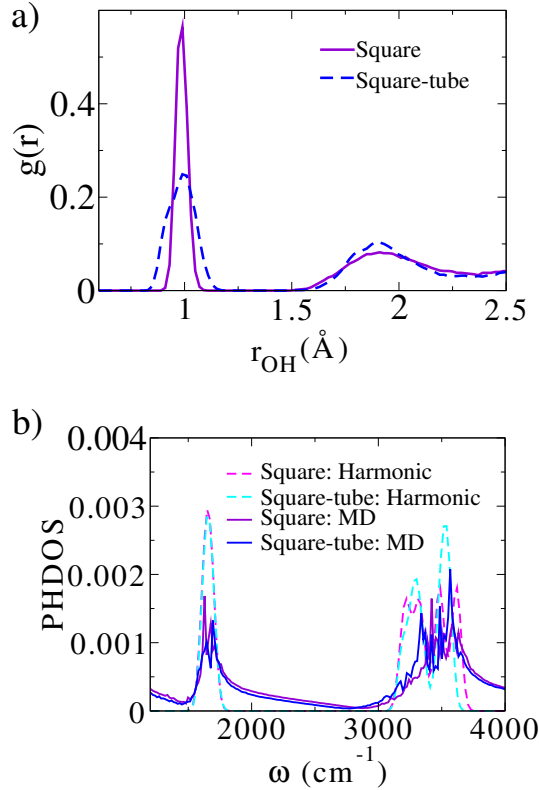}
\end{center}
\caption{
(a) Distribution of OH bond length for the square and square-tube structures as obtained from \textit{ab initio} molecular
dynamics simulations at 300 K.
(b) Phonon density of states for the square and square-tube structures.
Dashed lines: harmonic phonon density of states using the finite displacement method.
The calculation was performed 
with a $2 \times 2 \times 1$ supercell and a $8 \times 8 \times 1$ grid 
in the phonon zone using the Phonopy package \cite{phonopy}.
Solid lines: phonon density of states calculated from \textit{ab initio} molecular
dynamics at 300 K.
}
\label{figure6}
\end{figure}

\textbf{Finite temperatures:}
Considering that the experiments from which square ice has been suggested have been carried out at room temperature \cite{algara-siller_square_2015},
it is also important to evaluate the influence of finite temperature.
Therefore, the free energy instead of the enthalpy should be calculated.
As a first step the phonon free energy has been calculated at the harmonic level as,
$E_\text{vib}=\int \hbar \omega [\frac{1}{2}+\frac{1}{\text{exp}(\hbar \omega / k_{B}T)-1}] g(\omega) d\omega$,
where $g(\omega)$ is the phonon density of states (Fig. \ref{figure6}), $k_{B}$ the Boltzmann constant and $T$ the temperature.
At 300 K the phonon free energy of the square structure is
6 meV/$\text{H}_\text{2}\text{O}$ less than that of the square-tube structure.
Thus finite temperature effects increase the stability of the square structure,
and even though the energy difference between the square and square-tube structures is very small,
the square-tube structure remains marginally more stable.

\textbf{Role of anharmonic effects:}
It is also possible that anharmonic phonon effects at finite temperature
alter the stability of ice structures. 
Recently Engel \textit{et al.} found that it can result in a difference
of 6.5$\pm$3.1 meV/$\text{H}_\text{2}\text{O}$ for bulk ice I\textit{h} and I\textit{c}, which is
large enough to explain the difference between harmonic theories 
and experiments \cite{engel_anharmonic_2015}.
They also showed that the anharmonic contribution mainly comes from 
the vibrational modes at high frequencies that are related to the motion of hydrogen atoms.
Therefore, we carried out \textit{ab initio} molecular dynamics simulations at 300 K
for both the square-tube and square structures. 
Instead of accurately computing the challenging anharmonic free energy, here we try to estimate the
influence of anharmonic effects qualitatively from these molecular dynamics simulations.
Fig. \ref{figure6}a shows the distribution of OH bond lengths, which reflects the different
types of hydrogen bonds in the two structures.
The phonon density of states of these high frequency modes calculated from these trajectories are
shown in Fig. \ref{figure6}b, where no significant difference is observed.
We can also do the integration using the harmonic phonon free energy equation with these phonon density of states, which 
further shows that they are very similar as the integration gives a vanishing difference.
Therefore, these simulations indicate that the anharmonic contribution is unlikely to alter the relative stability of
the square and the square-tube structures.

\textbf{Role of hydrogen ordering:}
In the discussion above the lowest enthalpy square and square-tube structures are compared.
It is also important to consider how our conclusions at finite temperatures would
be affected by the 
configurational entropy of hydrogen ordering (different distributions of hydrogen atoms 
within the same lattice of oxygen atoms), which is important in ice \cite{salzmann_preparation_2006, salzmann_jcp_2016}.
A transition from the square-tube structure to the HCP structure driven by configurational entropy 
has been predicted in Ref. \onlinecite{corsetti_enhanced_2016}.
Here the possible configurations of different hydrogen distributions
of the square and the square-tube structure are counted
and the configurational entropy is estimated as $S=k_{B}ln\Omega$, where $\Omega$ is the number of
possible states. 
Following the Bernal-Fowler and Pauling ice rules \cite{Bernal_1933, Pauling_1935},
we impose three rules to the square phase: 
(i) hydrogen atoms locate in or between the two layers; 
(ii) half of the water molecules only bond to neighbors within the same layer,
the other half form one hydrogen bond within the layer and one hydrogen bond with the other layer;
and 
(iii) between each oxygen atom pair there should be no more than one hydrogen atom.
This leads to $\Omega_{square}=[C_4^2 \times (C_4^1)^8]^{1/8} \times (3/4)^2$.
For the square-tube phase the rules are: 
(i) no hydrogen atoms are allowed outside of the tube;
(ii) cells must be connected along the tube by more than (1/2) hydrogen bonds per water molecule;
and (iii) between each oxygen atom pair there should be no more than one hydrogen atom.
This leads to $\Omega_{square-tube}=(C_8^4)^{3/8} \times (3/4)^2$.
The contribution to the free energy difference between these
two phases is $\Delta F \approx 0.02 k_{B}T$ ($<$ 1 meV/$\text{H}_\text{2}\text{O}$),
which is negligible compared to the difference between the square and square-tube structures.
Although this estimate is crude and is based on the assumption that all possible hydrogen ordered states are degenerate, it suggests that configurational entropy
effects are unlikely to alter the relative stability of the two structures significantly.

\footnotetext[2]{
The lateral lattice constant and pressure
of the square ice calculated are in agreement with the TEM measurements of lattice constant
and pressure estimate.
The lattice constant calculated using the rPW86-vdW2 functional at 1 GPa is 2.82 \AA, and 
likewise a 2.83 \AA~lattice constant
predicts a pressure of 0.91 GPa. 
There are a few aspects that need to be noted:
(i) Thermal expansion is not considered;
(ii) The estimated pressure in the experiments is based on a simple model whose accuracy is not clear;
(iii) The relation between the lattice constant and pressure depends on the functional used, for instance
the optPBE-vdW functional gives a lateral pressure of 0.65 GPa with a lattice constant of 2.83 \AA;
(iv) The relation between the lattice constant and pressure depends to some extent on the definition of
lateral pressure, or in other words, the definition of the layer height. 
For double-layer ice our 
definition might underestimate the pressure, whereas the upper limit of pressure is about 50\% larger
as noted above. 
}

In this section we have addressed issues including 
the sensitivity of the results to the XC functional, ZPE, 
harmonic and anharmonic phonon free energies, and 
configurational entropy.
There are other issues not addressed which might also be important.
For instance, the presence of real graphene instead of a confining potential,
the finite size of the ice structures observed in the experiments 
and the role of edges.
We believe these effects will not affect our conclusions significantly because:
(i) The graphene is incommensurate with the ice and the interaction between graphene and water molecules is
not very sensitive to the orientation and lateral position of water molecules, which we have
shown in our previous study \cite{chen_two_2016}.
(ii) More hydrogen bonds break at the edge of the square structure than the square-tube structure.
Thus, on the basis of our calculations, we conclude that double-layer square ice is a metastable
phase across a wide range of confinement widths and pressures.
%From a different perspective from Zhou that NaCl generates the same experimental data as square 
%ice \cite{zhou_observation_2015}, 
%our results raise a question to the stability of double-layer square ice.
Interestingly, our previous work found that monolayer square ice is a stable phase
\cite{chen_two_2016, chen_qmc}, which lends support to
the experimental observations of square ice \cite{algara-siller_square_2015}.
Therefore, although we find double-layer square ice is metastable, 
it is possible that it forms as the growth of a metastable double-layer
phase assisted by the presence of a stable monolayer.
The influence of crystal growth kinetics and nucleation of square ice
under confinement would therefore be an interesting issue to
explore in future work.
Experimentally, some well controlled annealing experiments could also
be informative and help to establish if square ice is metastable. 
Zhou \textit{et al.} questioned the observations of square ice from a different perspective 
suggesting that the
confined material is layers of NaCl instead of ice \cite{zhou_observation_2015, algara-siller_algara-siller_2015, wang_wang_2015}.
Such a concern also calls for further experimental and theoretical investigations of both confined water and confined NaCl.
%Our study also inspires further studies to substantiate the stability of confined double-layer ice
%using methods at a higher level, such as quantum Monte Carlo.

%\textcolor{red}{summary and outlook}

\section{summary}

In summary, on the basis of DFT calculations a phase diagram of double-layer ice as a function of
pressure and confinement width at 0 K is predicted.
Complementary to our previous studies on monolayer ice \cite{chen_two_2016, chen_qmc}, 
this study shows an interesting change of the phase diagram
from monolayer to double-layer ice in confinement due to the presence of
interlayer hydrogen bonds.
For monolayer ice, a pentagonal and a square structure appear in the GPa regime,
whereas for double-layer ice the pentagonal and square structures are metastable and 
in the GPa regime double-layer ice favours HCP and square-tube structures.
Beyond the general insights on the phase diagram of confined ice,
we have performed extensive calculations focusing on the double-layer square ice, which
has been a matter of debate recently between experiment and theory.
Our results suggest that the double-layer square ice is a metastable phase.

\section*{Acknowledgements}
J.C. and
A.M. are supported by the European Research Council
under the European Union's Seventh Framework Programme
(FP/2007-2013) / ERC Grant Agreement number
616121 (HeteroIce project). 
A.M. and C.J.P. are supported by the Royal Society
through a Royal Society Wolfson Research Merit Award.
C.J.P. and G.S. are also supported by EPSRC Grants No. EP/G007489/2 and No. EP/J010863/2.
C.G.S is supported by the Royal Society (UF100144).
We are also grateful for computational resources to the
London Centre for Nanotechnology, UCL Research Computing,
and to the UKCP consortium (EP/ F036884/1) for access
to Archer.

\bibliography{ref}

\newpage
\setcounter{section}{0}
\renewcommand{\thesection}{S\arabic{section}}%
\setcounter{table}{0}
\renewcommand{\thetable}{S\arabic{table}}%
\setcounter{figure}{0}
\renewcommand{\thefigure}{S\arabic{figure}}%
%\section*{Supporting Information}
\section*{Supplemental Material}

\noindent

This supplemental material contains:
the confining potential at different widths (Fig.~S1);
additional enthalpy data and structures in Fig.~S\ref{fig_s2} and Fig.~S3;
and structure files for phases reported in the main text.

\begin{figure}[h!]
\begin{center}
\includegraphics[width=5in]{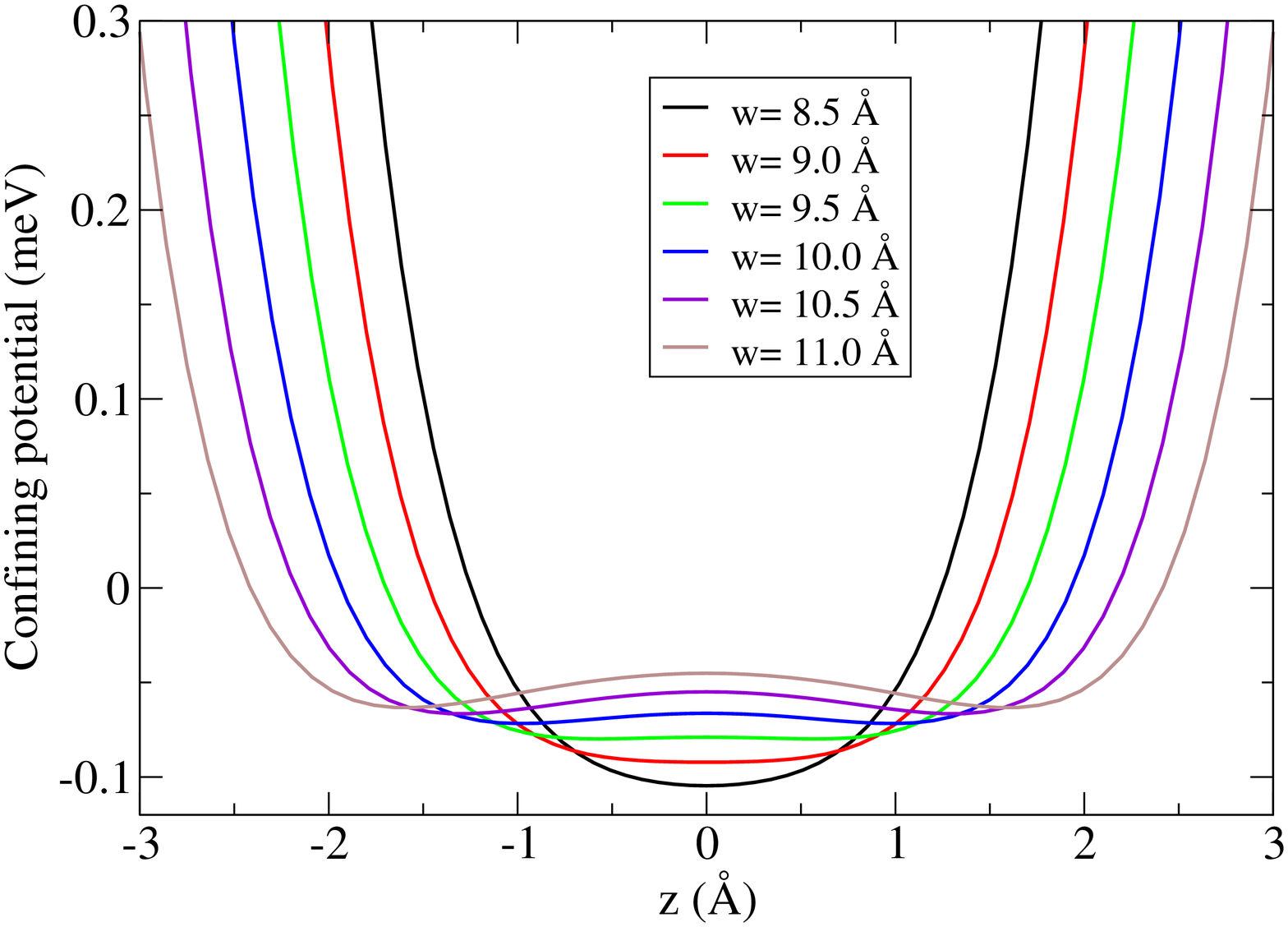}
\end{center}
\caption{
Confining potential at different widths.
}
\label{fig_s1}
\end{figure}

\begin{figure}[h!]
\begin{center}
\includegraphics[width=5in]{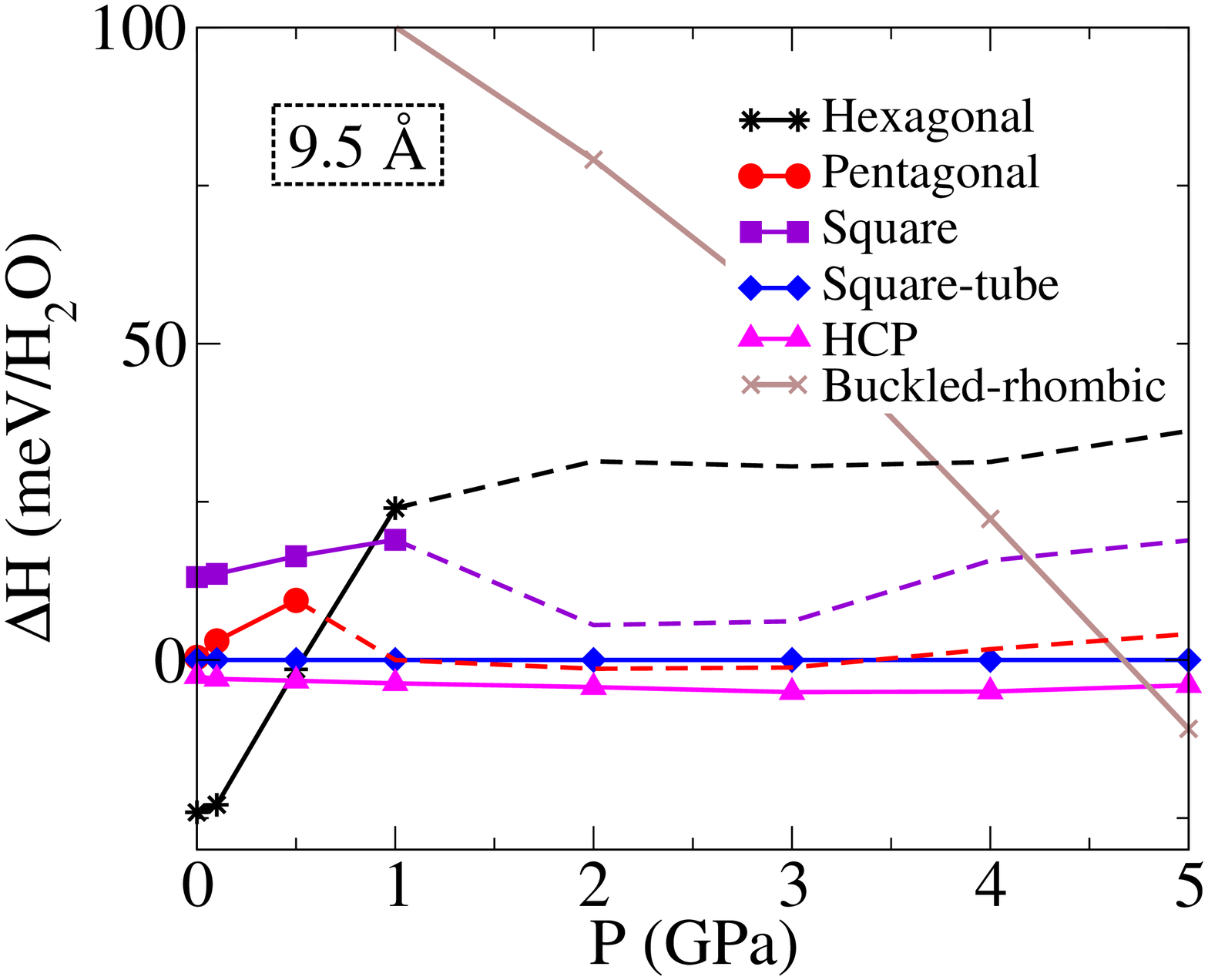}
\end{center}
\caption{
Enthalpies (relative to ST) of the double-layer ice structures 
as a function of lateral pressure
at a confinement width of 9.5~\AA.
Solid lines with filled symbols are the same as Fig. 3(c).
Dashed lines show the enthalpy of the structures obtained via structure optimisation starting with the hexagonal, pentagonal, and square phases at high pressure.
The corresponding structures are shown in Fig.~S2.
The data points of the dashed lines are at 1, 2, 3, 4, and 5 GPa.
We note that the small regime where the red dashed line has the lowest enthalpy does not indicate that we should have a new phase there. The red dashed line is the ST structure, with a larger unit cell and thus a slightly different hydrogen ordering. 
}
\label{fig_s2}
\end{figure}

\begin{figure}[h!]
\begin{center}
\includegraphics[width=5in]{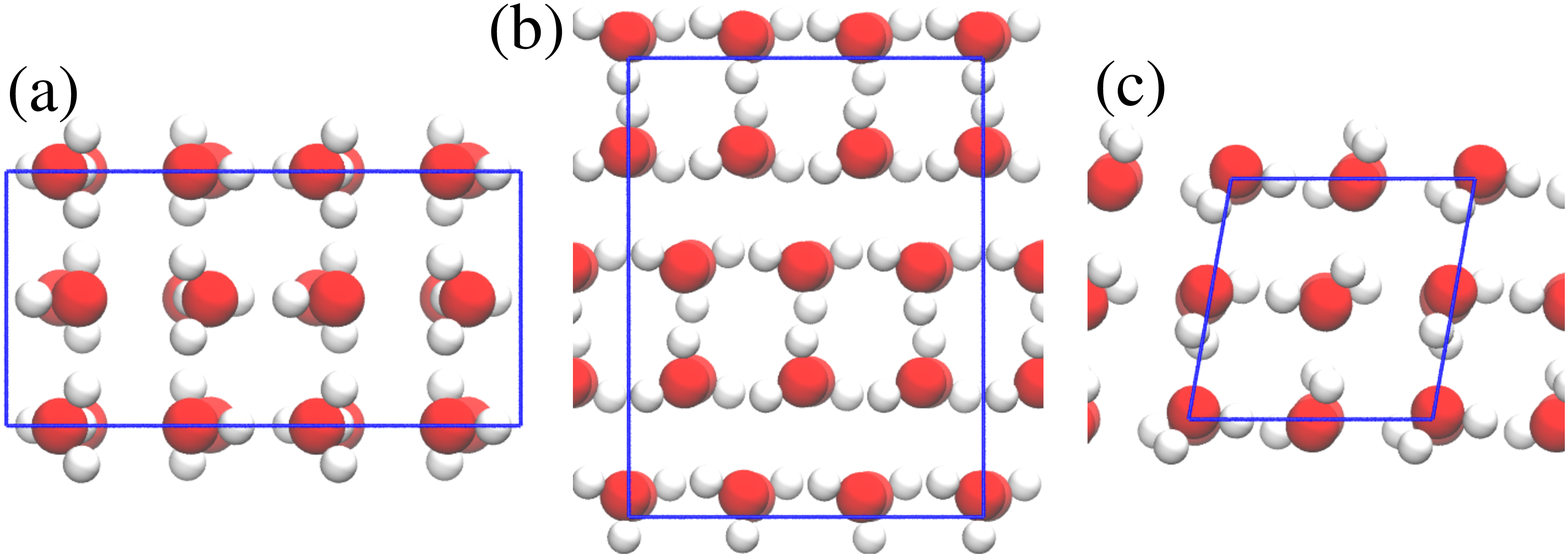}
\end{center}
\caption{
The structures corresponds to the dashed lines in Fig. ~S1. 
(a) A square structure, with a different unit cell and hydrogen ordering, that comes from optimising the hexagonal phase at 2 GPa.
(b) A ST structure, with a different unit cell and hydrogen ordering, that comes from optimising the pentagonal phase at 1 GPa.
(c) A tilted square structure that comes from optimising the square phase at 2 GPa.
Same structures were seen at higher pressures up to 5 GPa.
}
\label{fig_s3}
\end{figure}

%\newpage

%\bibliography{ref}
%\bibliographystyle{unsrt}

%\include{manuscript.bbl}

\end{document}